\documentclass{ws-ijmpd}
\usepackage[super,compress]{cite}
\usepackage{float}
\usepackage{color,array}
\usepackage{xcolor}
\usepackage{tabulary,booktabs,multirow}
\usepackage{nohyperref}  
\usepackage{url}  

\begin{document}

\markboth{Bojnordi \& Rahvar}
{Close stellar encounters kicking planets out of HZ}

%
\catchline{}{}{}{}{}
%

\title{\textbf{CLOSE STELLAR ENCOUNTERS KICKING PLANETS OUT OF HABITABLE ZONE IN VARIOUS STELLAR ENVIRONMENTS}}

\author{BEHZAD BOJNORDI ARBAB}

\address{Department of Physics, Sharif University of Technology\\
P.O. Box 11155-9161, Tehran, Iran\\
bojnoordi\_b\_@physics.sharif.edu\\
behzadbojnordi@gmail.com}

\author{SOHRAB RAHVAR}

\address{Department of Physics, Sharif University of Technology\\
P.O. Box 11155-9161, Tehran, Iran\\
rahvar@sharif.edu}

\maketitle

\begin{history}
\end{history}

\begin{abstract}
Continuous habitability of a planet is a critical condition for advanced forms of life to appear, but it can be endangered by astronomical events such as stellar encounters. The purpose of this study is to analyze close stellar encounters able to change planetary orbits initially in circumstellar habitable zones and to investigate the expected encounter rates in a variety of stellar environments. Using gravitational simulations for three-body systems, this study analyzed the dependencies of encounter impact-parameters with kinematic, geometric, and habitability parameters of the system. We also used kinematic properties of various stellar regions and estimated encounter rates of the events. The expected number of threatening stellar encounters in the Solar neighborhood is $\approx4.3\times10^{-4}$ in 4 billion years, while for the Galactic bulge environment, we expect approximately 5.5 times the value. The encounter rates for other stellar environments are calculated and spheroidal dwarf galaxies and globular clusters encounter rates are estimated. The results show that in contrast with the solar neighborhood, close stellar encounters can play a significant role in the expected number of planets with continuous habitability in dense stellar environments. Another notable result shows that threatening stellar encounter rate follows the number density of stars, and is not strongly dependent of the region's velocity dispersion. Further investigations are needed to study long-term multiple planetary systems and how they can change the overall expected value of continuously habitable planets.
\end{abstract}



\section{Introduction}	

The search for other planets harboring life is one of the most ambitious projects in astronomy. As thousands of confirmed exoplanets are discovered in recent years ,\cite{NASA-ea} the interest in finding habitable planets and eventually life-harboring planets increases.
One of the main environments to search for life is the circumstellar habitable zone (CHZ or generally HZ, as used in this paper); the region around stars where a rocky planet in possession of an atmosphere can support liquid water. For a planet located in this region, life can potentially emerge, sustain and evolve \cite{Huang_The_1959, Huang_The_1960, Kasting1993, Kopparapu_HABITABLE_2013}.

However, simply finding a planet in HZ is not a promise for life. Assuming emergence of life at a moment in the history of the planet, continuous habitability is required for a sustained ecosystem, and there might be catastrophic events threatening it. Various events originated from the planet can cause mass extinctions. For example, volcanic activities and eruptions causing catastrophic climate change \cite{Wignall2001}, as several mass-extinctions have been characterized to be coinciding with mass-volcanism events ,\cite{Alvarez2003} with possible astronomical stimulations like stellar encounters \cite{Pandey1987, Abbott2002}.

Further than being a stimulant for terrestrial catastrophes, stellar encounters play a major role in \textit{extraterrestrial} catastrophes and stellar population evolutions. Stellar encounters have been a very important phenomenon in relatively dense stellar environments. In recent years, many astronomers studied stellar encounters for different reasons and in various scenarios, including planetary ejections and rogue planets, interactions with circumstellar protoplanetary disks, and perturbations on planetary orbits or the Oort cloud (as stated above). As for one of the first attempts for calculating the effects of stellar encounters, Ref.~\refcite{Lyttleton1965} calculated the changes in planetary orbital parameters from a passing star by the method of variation of elements, and also took into account the cumulative effects of multiple encounters. One of the early motivations to study such encounters came from primitive predictions (e.g. Ref.~\refcite{Rasio1996}) and observations (e.g. Ref.~\refcite{Luhman2005}) of rogue planets. In an attempt to calculate the number of rogue planets in star clusters, Ref.~\refcite{Hurley2002} performed N-body simulations for the M22 cluster, and taking into account the interactions between a set of Jupiter-mass planets and the nearby cluster stars, calculated the percentage of planets ejected by the encounters, and the fraction that leave the cluster afterward. Ref.~\refcite{Veras2012} also showed that the planet-planet interactions can not explain the majority of rogue planets alone, so the stellar encounters can be responsible for the remaining bulk. 

In the early formation-period of a planetary system, stellar encounters can be critical in the properties of later protoplanetary disks and the planetary system. Ref.~\refcite{Kobayashi2001} investigated the changes in eccentricity, inclination, and longitude alignments of the protoplanets in inner and outer regions of protoplanetary disks, caused by an encounter event.
Stellar encounters can also change the density profile and shrink the size of protoplanetary disks. Ref.~\refcite{vincke2015} investigated the truncation of the disks in different cluster populations, showing that in birth clusters like Orion Nebula Cluster, stellar encounters play an important role in shaping the protoplanetary disks. We can also expect star capture, \cite{Clarke_Star_1991} ejection of disk matter, or an angular momentum transfer to the star. \cite{Hall_Energetics_1996, Ostriker_Capture_1994} Stellar encounters in young clusters can strip stars of debris disks,\cite{Lestrade2011} change planets' orbital eccentricity, and capture or send planets out of the planetary system. \cite{Li2015, Malmberg_The_2011, Jimenez-Torres2011} However, the consequences of stellar encounters can be altered or canceled through dynamical processes of later planetary formation stages, as Ref.~\refcite{Marzari2013} and \refcite{Picogna2014} showed that the circumstellar disk can erase the stellar encounter effects on protoplanets eccentricities in $ \approx10$~{kyr} during the stay in the birth cluster, but the semi-major axes do not revert to the initial sizes. As a result of the encounter effects on planetary systems, Ref.~\refcite{Zwart_The_2015} distinguished the circumstellar regions in open clusters based on gravitational effects from external and internal planetary system influences by defining Parking Zone and Frozen zone. These zones specify the size of the circumstellar orbital space mainly uninfluenced by stellar encounters.
Furthermore, Incidents of such encounters in dense stellar clusters change the population of planets and play an important role in planetary formation procedures \cite{Jimenez-Torres2011, Fragner2009, Craig2013,  Malmberg_The_2011}.

 Stellar encounters can produce catastrophes in various ways. For instance, the encounters can increase major-impacts of asteroids or comets which can cause life-threatening cataclysms in a wide range of danger, from threatening many species, to wiping out life from the face of Earth \cite{Chapman1994}. Almost all long-period comets come from the Oort cloud \cite{Oort1950}, and due to perturbations, comets may inject into the inner solar system region. Ref.~\refcite{Rickman2008} studied the effect by modeling an Oort cloud and random stellar encounters to measure the impact of stellar and cometary orbital parameters on injections of comets into the inner solar system.
 The outer Oort cloud radius is comparable to the distance of nearby stars and as a result, the gravitational force of a "stellar encounter" can induce a shower of comets into the inner planetary system, increasing major-impact rates \cite{Hills1981, Matese2002, Wickramasinghe2008}. 
 Ref.~\refcite{Bobylev2017} studied the possible past and future stellar encounters from nearby stars that can produce the perturbations in the Oort cloud, and Ref.~\refcite{Bailer2017} utilized data from the first Gaia data release (GDR1) and other projects, and integrating the stellar motions through Galactic potential, identified close stellar encounters comparable to the Oort cloud. More recently, Ref.~\refcite{Rybizki2018} implemented almost the same method for the second Gaia data release (GDR2). 
 
 As a result of the the weak binding of the Oort cloud to the solar gravitation, other mass-encounters including the tidal force of the Galaxy and dark matter disk \cite{Rampino2015}, or giant molecular clouds (GMCs), can perturb Oort cloud enough to alter the rate of comets falling in the inner regions of the solar system. \cite{Mazeeva2004, Jakubik2008} Ref.~\refcite{Feng2014} extended the stellar and galactic encounter models to study the variations of long-period comets with regard to gravitational perturbation from galactic tide and stellar encounters.inspired by these results, Ref.~\refcite{Martinezbarbosa2016} took into account different galactic orbits of the solar system, and how it can change the rate of stellar encounters that can effectively interact with the Oort cloud, considering stellar encounters within $10^5AU$ from the sun.


The Oort cloud disturbance can increase the rate of dangerous cometary impacts, but stellar encounters can introduce other phenomena which can devoid the planet of habitability conditions. 
As stated above, stellar encounters can eject planets from the planetary system, and therefore the circumstellar habitability conditions don't apply to planet anymore. \cite{Lyttleton1965, Veras2012, Zwart_The_2015, Cai2017, VanElteren2019a} The main motivation for this study is to examine the less discussed phenomena in the stellar encounters; ones which can change the orbit of the planet, enough to eradicate habitability from the planet, but not necessarily ejecting the planet out of the system. In contrast to the Oort cloud disturbing encounters, as a result of smaller planetary orbits compared to comet orbits, this type of encounters occur with much smaller impact parameters, and lower expected encounter rates.
Consequently, we specifically focus on the encounters which endanger the persistence of planetary orbits in the "habitable zone". The HZ discussed is from the conventional definitions for Earth-like planets (in contrast to satellites of gaseous planets, or habitability beneath the surface). Here we don't study the possibility that the planet changes back to the HZ in short time-scales. 

Also, we used the time-scale of life on Earth to study the chances of stellar encounters during this time, that can endanger the continuous habitability of the planet. We present results in the solar neighborhood, Galactic bulge, birth clusters, globular clusters, and spheroidal dwarf galaxies. The information on the abundance of such encounters can identify environments that the effects of this phenomena should be taken into account and how it can change our expectations on the number of planets harboring life in different regions.

Note that the changes in other orbits in the planetary system and the inter-planetary gravitational effects are yet to be studied. The results of this study should be considered as the direct gravitational effect of the stellar encounter. The addition of other planets and taking into account the interplanetary interactions make the simulations much more expensive. The present setting enables us to study and compare stellar environments with respect to significance of this catastrophe. 

This paper is organized as follows. In section (\ref{methods}), we express the methods used for simulating the close stellar encounters, defining the critical impact-parameter and cross-section, threatening impact parameter and cross-section, and threatening encounter rate.  In section (\ref{Results}), we study the correlations between the critical impact parameter, the initial parameters of the encountering star, and the habitable zone width, and finally, calculate the critical encounter rates for the solar system neighborhood and the Milky Way bulge. The discussions and conclusion are given in sections \ref{Discussions} and \ref{Conclusions}.

\section{Methods and Data}
\label{methods}
In this section, we begin by introducing the close stellar encounter and associated parameters. Also, we discuss the rate of encounters that can deviate the planet's orbit from the habitable zone.\subsection{Methods}\label{Methods1}For simplicity, we assume a gravitational system consisting of three bodies: (i) the parent star, (ii) the orbiting planet around the parent star, and (iii) the encountering star. We simplify the problem by assuming that the target star has one planet, which together with the encountering star they create a three-body system. For the encountering stars, we use the kinematic and the stellar population of stars in the Galaxy while we adapt the target star as a solar-type star and the planet as an Earth-type planet \cite{Garcia-Sanchez2001}. Here for simplicity, we ignore binary and multiple stars.

\begin{figure}
	\centerline{\psfig{file=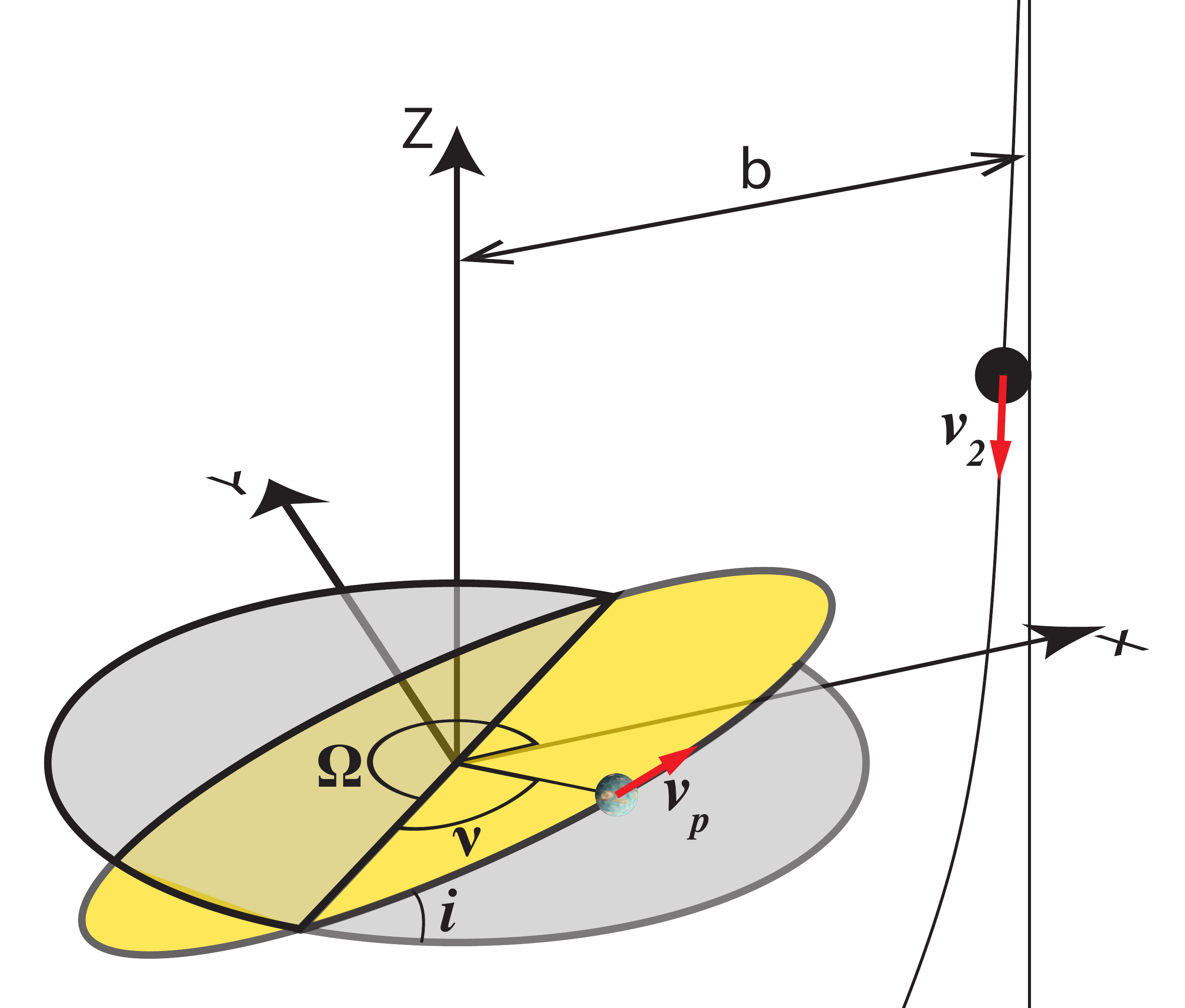,width=0.6\linewidth}}
	\caption{The orbital parameters of encounters are as follows: $b$ is the impact parameter of the secondary star from the parent star, $v_2$ is the initial velocity of the secondary star, $v_p$ is the initial orbital velocity of the planet, $\Omega$ is the longitude of the ascending node, $i$ is the inclination of the planetary orbit from the (x-y) plane and $\nu$ is the initial anomaly of the planets measured counterclockwise from the ascending node. 
	} 
	\label{fig:orbit}
\end{figure}

For the numerical calculation, we use the gravitational integrations for the stellar encounter by the REBOUND code\footnote{https://rebound.readthedocs.io}
\cite{Rein_REBOUND_2012} and the IAS15 algorithm \cite{Rein2014a}. REBOUND is a 15th-order gravitational dynamics integrator and optimized for integrating close-encounter schemes. The encountering parameters in our study are (i) the velocity and mass of the encountering star (${v_2}$ and ${M_2}$), (ii) the impact parameter of the encountering star from the primary star ($b$), (iii) two angular parameters indicating the orientation of the planetary orbital plane from the encountering star (i.e., $\Omega$ and $i$ in Figure \ref{fig:orbit}), and (iv) the initial anomaly of the planet (i.e.,$\nu$ in Figure \ref{fig:orbit}).
We also include a parameter indicating the width of the habitable zone ($W_{hz}$). 
Here, we adopt the mass of the parent star to be one solar mass and the planet to be a test particle.

We place the origin of the Cartesian coordinate system on the position of the parent star.
The secondary star is initially at the position of $ (x,y,z)=(b,0,50{AU})$, along the $z$ axis in this coordinate.  
Two angular parameters are required to specify the initial alignment of planet orbit in this coordinate system; the inclination angle of "$i$" from the $xy$ plane and the longitude of the ascending node ($\Omega$) of the planetary orbit which is measured from the $x$-axis.
Also, we measure the anomaly of the planet (i.e., $\nu$) from the ascending node, counterclockwise on the planet's orbital plane.

In the simulations, we examine the mentioned parameter space with the following range of $\Omega{\in}[0,2\pi]$, $\nu{\in}[0,2\pi]$, $i{\in}[-\pi/2,\pi/2]$, $b{\in} (a_p{\times}[0.1,100])$, where $a_p$ is the planet orbital radius. The parameters $\Omega, \nu$, and $b$ are chosen with uniform distribution functions. Due to choosing the orbital parameters from the Euler angles, the initial planetary positions are biased to be near the XY plane. As a result, we choose inclination parameter ($i$) from a uniform distribution of $\cos (i) \in [-1,1]$. We discuss the distribution functions of the encountering star velocity ($v_2$) and the mass of the secondary star ($M_2$) in the next section. For each run of the simulation, the initial parameters are selected from the mentioned $6$-dimension space and finish the calculation of the dynamics when the encountering star reaches the distance of $z = - 50$~{AU} far from the primary star. After the encounter, we calculate the orbital parameters of the planet. 

\subsection{habitable-zone expelling/safe encounters and critical impact parameter}
\begin{figure}
	\centering
	\centerline{\psfig{file=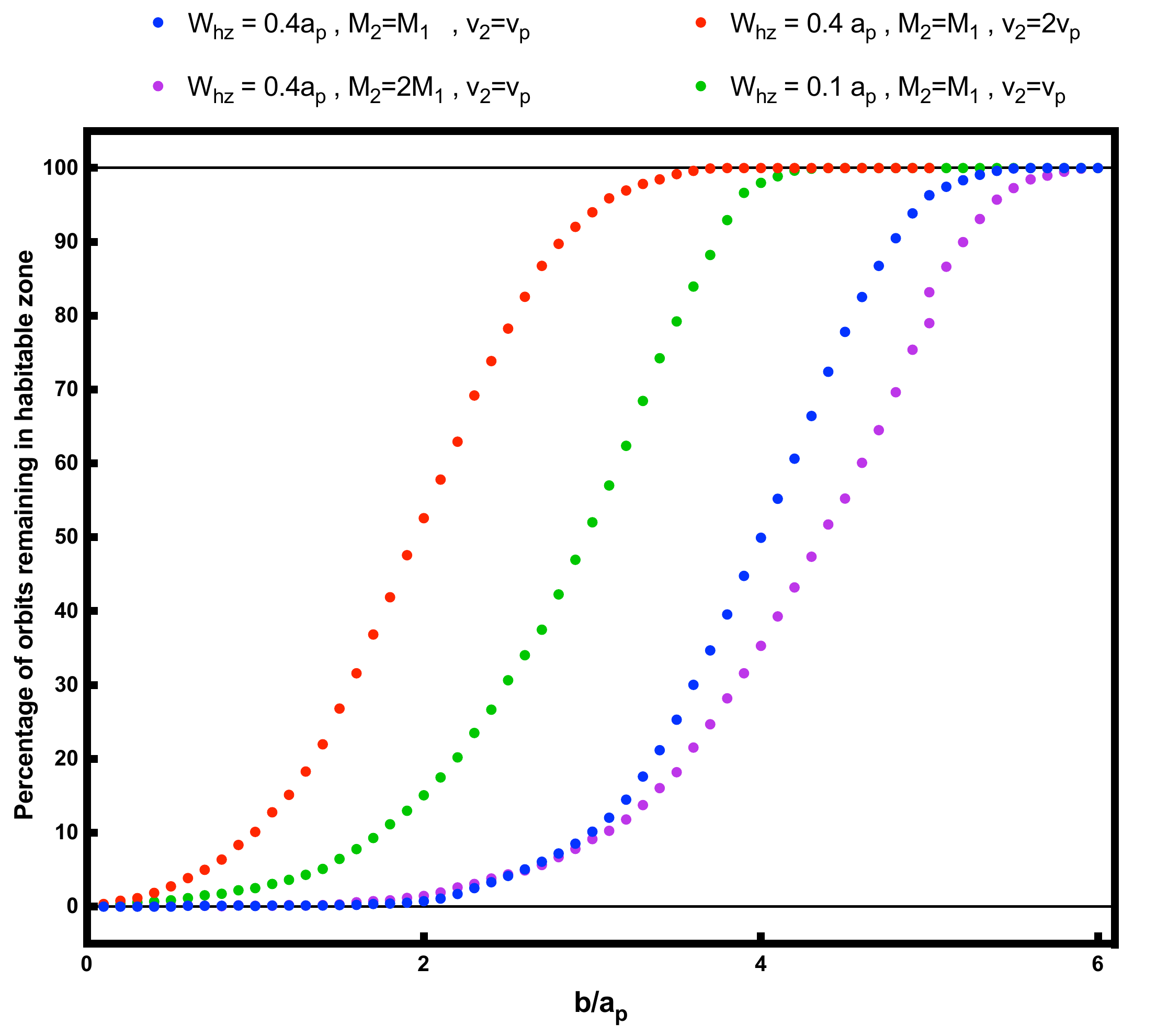,width=0.6\linewidth}}
	\caption{Percentage of orbits remaining in the habitable zone for different initial conditions.
	}
	\label{fig:chances}
\end{figure}

An event is a "habitable-zone expelling encounter" if perturbations from the encountering star push the planet out of the habitable zone via the inner or outer boundaries of HZ, even if most of the planet is still in the HZ area. In contrast, it is a "safe encounter" if the planet's orbit stays entirely in the HZ after the encounter. If not stated otherwise, the habitable zone boundaries used in this study is Kasting\cite{Kasting1993} conservative instantaneous HZ ( 0.95-1.37 au). In each encounter, the total numerical relative energy error is less than $\Delta E/E <10^{-14}$, and the total numerical relative angular momentum error is less than $\Delta L/L< 10^{-15}$. Note that the habitable-zone expelling encounter, by definition, is just about a specific impact parameter and does not say anything about smaller impact parameters, but evident from Figure \ref{fig:chances}, one can conclude that in larger impact parameters, the encounters will be safe ones.

\begin{figure}
	\centering 
	\centerline{\psfig{file=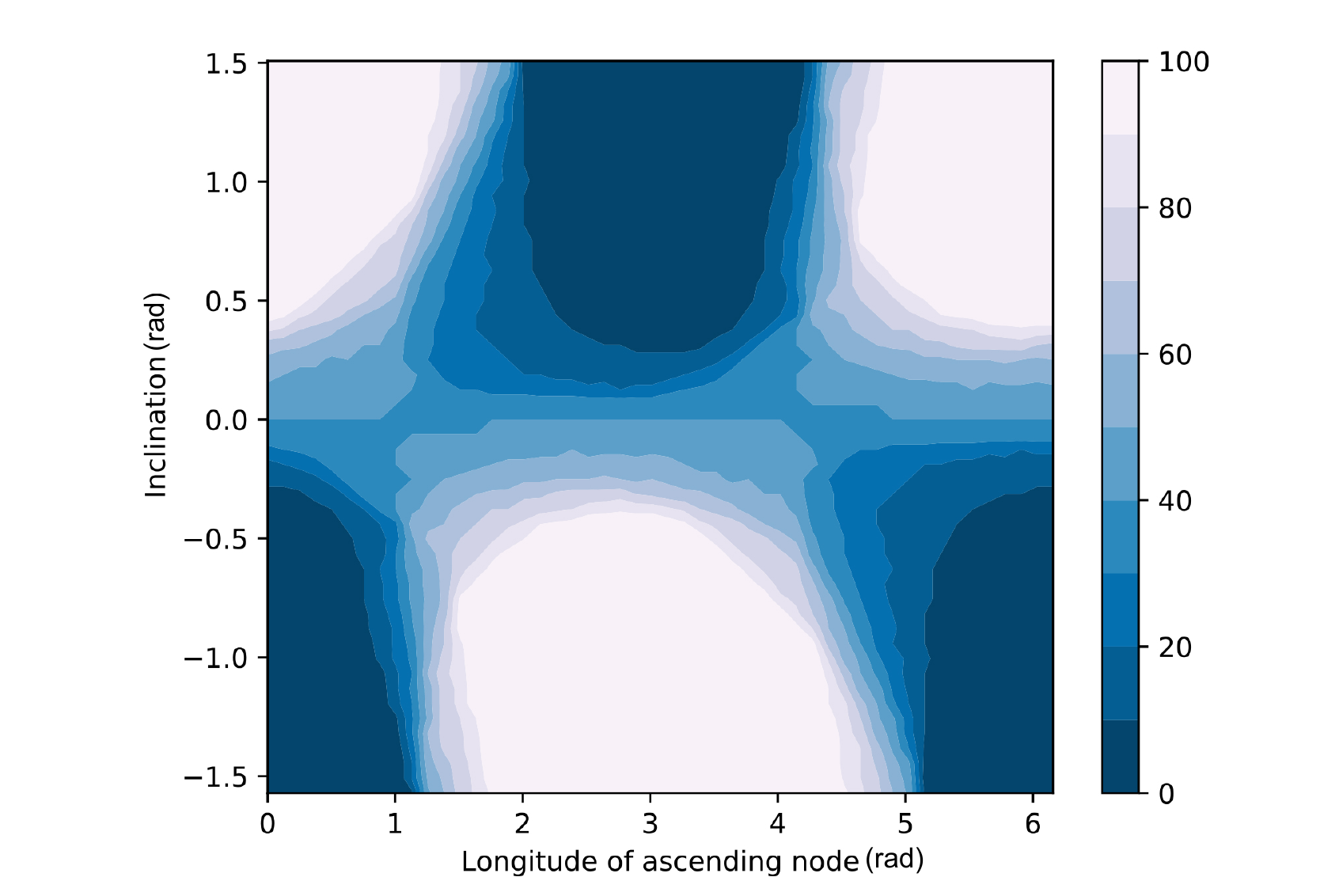,width=0.6\linewidth}}
	\caption{Probability function of the planetary orbit that remain in the habitable zone in percent. Here we adapt $M_2 = 1 M_\odot$, $v_2 = v_p$ and $b = 4a_p$, the the primary system is identical to the sun-earth parameters, Kasting\cite{Kasting1993} conservative instantaneous HZ is used, and the percentage is calculated from a pool of initial anomaly of the planet. Both inclination and longitude of ascending node are measured in radians.
	}
	\label{fig:orientation}
\end{figure}

Figure \ref{fig:orientation} demonstrates an example from our simulation where we fix three parameters of $M_2$, $v_2$ and $b$. Here $v_p$ is the orbital velocity of the planet around the parent star. In this figure, we identify stellar encounters in terms of $\Omega$ versus $i$ integrated over all the values of anomalies. The mean percentage over the plot shows the probability function for the planet remaining in the habitable zone of the parent star ($P_{HZ}$). The probability function for the planet exiting the habitable zone is $1{-}P_{HZ}$.

In what follows, we classify the parameters of this problem into two groups as $(M_2, v_2, b, W_{hz})$ and the second group identifying the angular orientation of the planet as ($i$, $\Omega$ and $\nu$).
We will integrate over the parameters in the second group to provide the probability function of a planet remaining in the HZ  (i.e., $P_{HZ}$), in terms of the first group of parameters. 
Figure (\ref{fig:chances}) represents another example from our simulation where we calculate $P_{HZ}$ 
as a function of impact parameter for various sets of the mass and the velocity of encountering stars and the width of the habitable zone. As shown in Figure (\ref{fig:chances}), we define the "critical impact parameter" ($b_c$) as the smallest impact parameter where $P_{HZ}$ is $100\%$.

We study the relationship between the critical impact parameter ($b_c$) and physical parameters such as the width of the habitable zone boundary ($W_{hz}$), velocity ($v_2$), and mass ($M_2$) of the encountering star, integrating over the rest of the parameter space.  Here we simulate 8000 encounters for each set of physical parameters and scanning uniformly over $i$, $\Omega$, and $\nu$.

\subsection{Threatening encounter rate}
\label{methods2}
In order to measure the rates of encounters that are threatening to the habitability of the planet, we should use an average value of HZ expelling/safe probabilities over cross-section elements. The value requires a probability-threshold, which we choose to be $1\%$, and call it the Threatening cross-section ($\sigma_T$): "the cross-section around the primary star, where encounters within have $1\%$ probability for the planet to exit HZ (a HZ expelling encounter happens)".

To calculate the threatening cross section, we integrate in the impact-parameter space over probability function of HZ expelling encounters:


\begin{equation}
\label{Threatening_Impact_Parameter}
\frac{\int_{0}^{b_t} 2\pi b (1-P_{HZ})db}{\pi b_T^2} =1\%
\end{equation}

\begin{equation}
\label{Threatening_Cross_Section}
\sigma_T=\pi b_T^2
\end{equation}

We note that the cross-section is a function of velocity and the mass of the secondary star and the width of the habitable zone. We can calculate the "Threatening Encounter Rate" ($\Gamma_{T}$) as the rate of flybys within the $\sigma_T$, which can be evaluated by: \cite{Hut1985, Garcia-Sanchez2001, Li2015}

\begin{equation}\label{encounter_rate}
\Gamma_{T} = n_{\star}\langle \sigma_{T} \rangle_{(v)} \sqrt{v_{\odot}^2 + v_{\star}^2}~,
\end{equation}

where $v_{\odot}$ is the peculiar velocity of the sun in local standard of rest in the solar neighborhood \cite{Bobylev2014}, $v_{\star}$ is the velocity dispersion of stars, $n_{\star}$ is the number density of the stars in the environment, and $\langle \sigma_T \rangle_{(v)}$ is the averaged threatening cross-section over the stellar velocities function in the environment $f_{(v)}$. We assume a Maxwell Boltzmann velocity distributioin function, where:
\begin{equation}\label{Maxwell-Boltzmann}
f_{(v)}=\sqrt{\frac{2}{\pi}}\frac{v^2e^{-\frac{v^2}{2\alpha^2}}}{\alpha^3},
\end{equation}

where $\alpha$ is the scale parameter, which has the following relationship with the velocity dispersion:
\begin{equation}\label{velocity dispersion}
v_\star=\sqrt{\frac{\alpha^2(3\pi-8)}{\pi}},
\end{equation}
therefore, the probability density function with regard to the velocity dispersion is:

\begin{equation}\label{maxwell}
f_{(v)}=  \frac{\sqrt{2(3\pi-8)^3}}{\pi^2 v_\star^3} v^2 e^{-\frac{(3\pi-8) v^2}{2\pi v_\star^2}}~.
\end{equation}

As a result, the threatening encounter rate ($\Gamma_{T}$) is specific for each stellar mass and corresponding velocity dispersion.

\section{Results}
\label{Results}

In this section, we present the underlying aims of the research. We study the relationship between the critical impact parameter with the physical parameters (i.e., the velocity and the mass) of the encountering star. Then, the relation of the critical impact parameter with the width of the habitable zone for various ranges is studied.
We also study the effect of the inclination angle of the orbital plane of the planet on HZ expelling encounters. 
This section concludes by presenting the results for Threatening cross-section and threatening encounter rates for various stellar environments. We also present the total threatening encounter rate and the probability of such encounters in the 4-billion years period.
We use Ref.~\refcite{Kasting1993} conservative definition of instantaneous HZ (and we call it Kasting HZ) throughout this paper unless indicated otherwise.

\subsection{Dangerous impact parameter relation with encounter parameters}
\subsubsection{The relation between the critical impact parameter and the velocity of the secondary star}

\begin{figure}
	\centering
	\centerline{\psfig{file=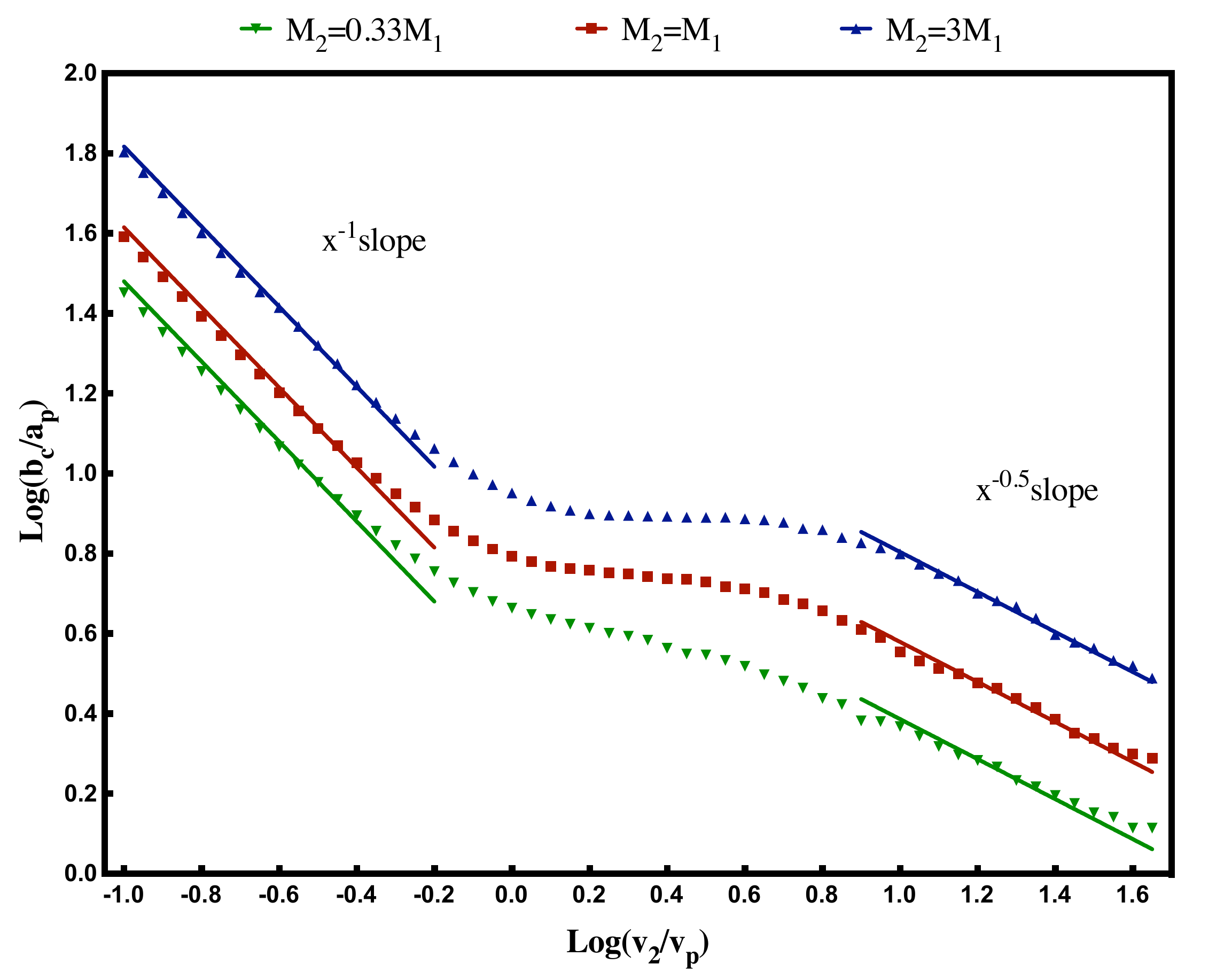,width=0.6\linewidth}}
	\caption{The critical impact parameter normalized planetary orbital radius, as a function of the relative velocity of the encountering star with respect to the velocity of the planet, both in logarithm scales. The three datasets are for  different secondary star masses;  
		$M_2=0.3M_1$, $M_2=M_1$, $M_2=3M_1$, and Kasting HZ is used. Logarithm bases are 10. The limits are chosen to demonstrate the changeover.
	}
	\label{fig:velocity}
\end{figure}

In order to investigate the relationship between the critical impact parameter and the initial velocity of the encountering star, we simulated systems with a broad range of velocities compare to the dispersion velocity of stars.  In Figure (\ref{fig:velocity}), we plot the critical impact parameter normalized to the orbital radius of the planet, as a function of the relative velocity of encountering star compared to the velocity of the planet. Here we have different scaling relations of the power of ${\sim}{-}1$ for the small and power of ${\sim}{-}0.5$ for the large velocities of the encountering stars. 

We can interpret Figure (\ref{fig:velocity}) from the rough analysis of the gravitational impact. Let us use the generic relation between the minimum encounter distance (i.e.,$r_{min}$) and impact parameter (i.e.,$b$)  from the three body problem\cite{Hills1984} as follows 
\begin{equation}\label{key1}
r_{min}=a_c \left( \left[ 1+ \left( \frac{b}{a_c} \right)^2 \right]^{1/2}-1 \right)~,
\end{equation}
where 
\begin{equation}\label{key2}
a_c = \frac{G(M_1+M_2)}{v_2^2},
\end{equation}\\
and $a_c$ is the accretion radius, parameterizing the gravitational focusing.

{\bf High velocity regime:}  In the case of a distant encounter 
and high relative velocities ($v_2\gg v_P$) the gravitational impulse difference($\Delta I$) between the primary star and the planet from the encountering star after the encounter (also known as heliocentric impulse) is proportional to \cite{Rickman1976}
\begin{equation}\label{imp}
\Delta I\propto\frac{M_2}{v_2 r_{min}^2}.
\end{equation}

For fast encountering stars, from equations (\ref{key1}) and (\ref{key2}), we can conclude that $r_{min}$ can be approximated with the impact parameter, $ b$. 
As a result of equation (\ref{imp}), and assuming that a specific minimum impulse is required to push the planet out of the habitable zone, we can conclude a proportionality between the dangerous impact parameter and the velocity of the fast encountering star:
\begin{equation}\label{key3}
b_{d} \propto v_2^{-0.5}~,
\end{equation}
which explains the plot slopes in the right part of Figure \ref{fig:velocity}. The high-velocity HZ expelling encounters of stellar masses lower than $1M_\odot$ occur with approximately asteroid belt distances.

{\bf Low velocity regime:} For a relatively slow encountering star (i.e. $v_2 < v_p$), the relative trajectory with respect to the primary star is almost parabolic, so the perihelion can define the trajectory. Low velocities make the accretion radius much larger than the dangerous impact parameter, therefore equation \ref{key1} becomes $r_{min} = \frac12 \frac{b^2}{a_c}$, and substituting $a_c$ from equation (\ref{key2}) results in:
\begin{equation}\label{key4}
b_{d} \approx  \frac{ \left[ 2r_{min} G (M_1 + M_2) \right]^{1/2}}{v_2}.
\end{equation}

As a result of the parabolic trajectory, we expect threatening encounters to happen in the same distances for each set of stellar masses, and therefor, $b_d\propto v_2^{-1}$.
This also means that we see the gravitational lensing in the encountering star flyby in low relative velocity, but not in high velocities. Such encounters happen in distances comparable to Uranus and Pluto orbital radii.

The critical impact parameter as a function of the low and high-velocity regimes of the secondary star 
is consistent with Figure  (\ref{fig:velocity}).

\subsubsection{Relation between the critical impact parameter and the mass of the secondary star}
\begin{figure}
	\centering
	\centerline{\psfig{file=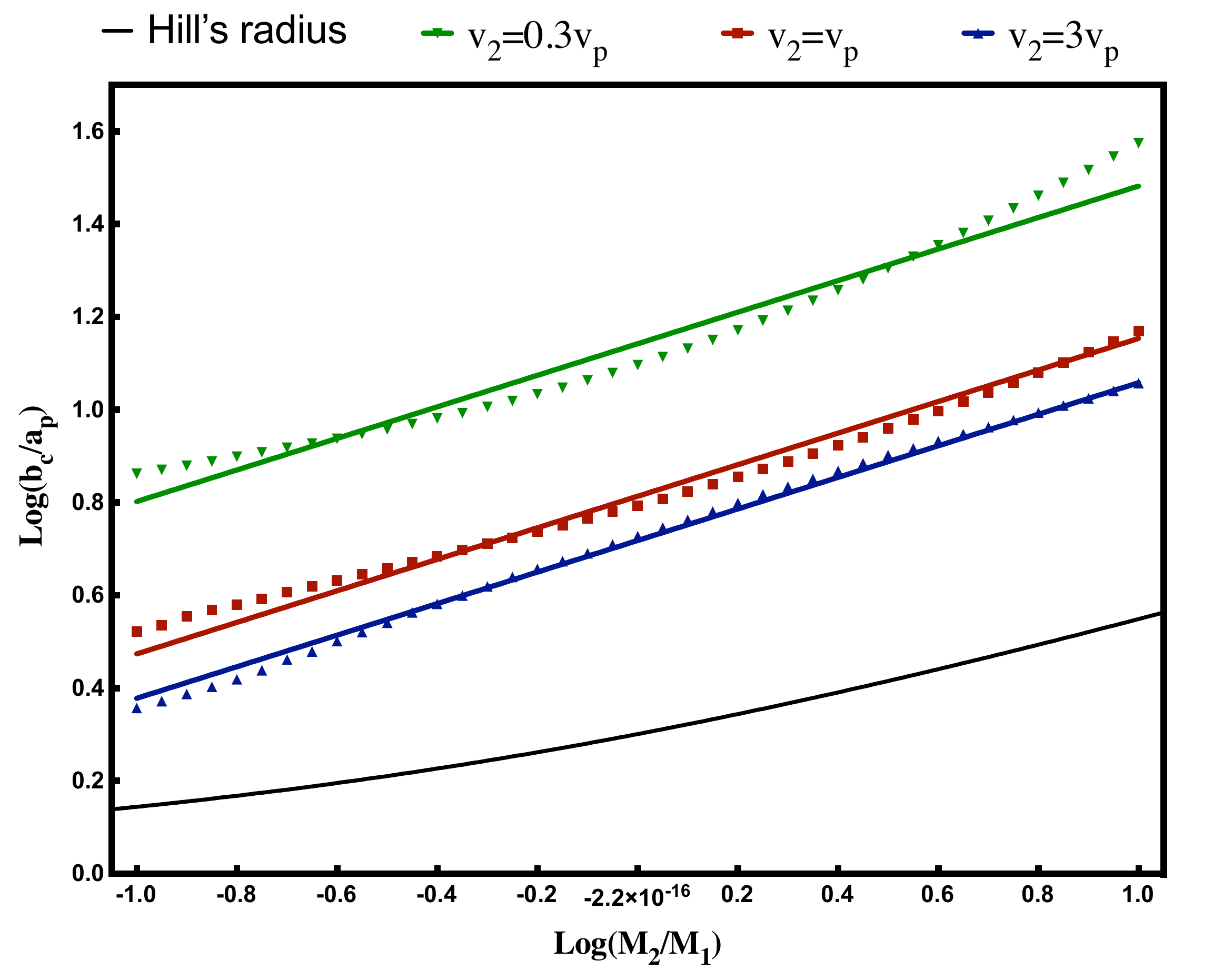,width=0.6\linewidth}}
	\caption{Log-Log plot of the critical impact parameter as a function of the mass of the secondary star, relative to the primary star for three different velocities. plotted for three different velocities: $v_2=0.3v_p$, $v_2=v_p$ and $v_2=3v_p$. The black curve is the Hill's radius without the negligible mass or orbital radius assumption. Note that here the $1/3$ slope occurs in higher mass ratios. 
		Kasting HZ boundaries are used.
	}
	\label{fig:mass}
\end{figure}

The relative mass of the secondary star to the primary star (i.e.,$M_2/M_1$) plays a significant role in the size of the critical impact parameter. We performed simulations for the relative mass in the range of $0.1<M_2/M_1<10$. The results of our simulation are represented in Figure (\ref{fig:mass}) for the impact parameter in terms of the relative masses for three different velocities of the secondary star. By increasing the mass of the secondary star, the tidal forces on the star-planet system increases, and the critical impact parameter occurs in the larger distances. 

Fitting the numerical results with a power-law function results in 
\begin{equation}\label{mass-relation}
\frac{b_{d}}{a_p} \propto (\frac{M_2}{M_1})^{0.34}~.
\end{equation}
This equation is almost similar to the Hill radius, $r_H\propto (M_2/M_1)^{1/3}$, which represents the unstable sphere around the primary star in the binary system. However, we note that in the Hill radius $M_2\ll M_1$, unlike the system in our concern. 
Also, $b_d$ is the impact parameter of the encounter, rather than the minimum distance of the stars.
In our notation, the mass-ratio is the inverse of mass ratios in Hill's problem, as $M_1$ is the and $M_2$ are Sun and the small planet, but here the orbit is around $M_1$, and the star with $M_2$ is passing. The orbital radii are inversed too, therefore the overall relations are still alike, but the $1/3$ slope in the log-log plot occurs in high mass-ratios.

\subsubsection{Relation between the critical impact parameter and width of the habitable zone}
\begin{figure}[pb]
	\centerline{\psfig{file=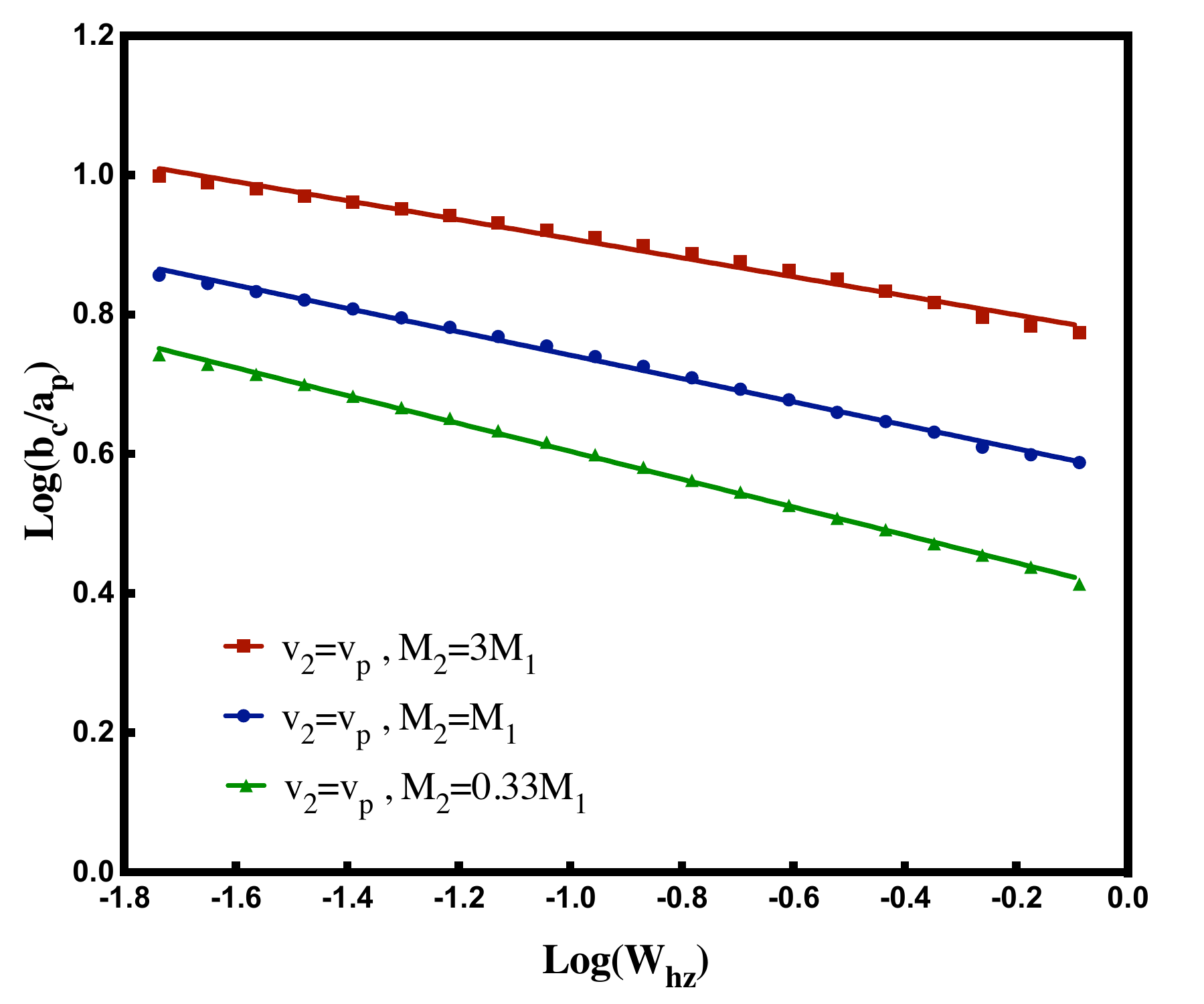,width=0.6\linewidth}}
	\vspace*{8pt}
\end{figure}
\begin{figure}[pb]
	\centerline{\psfig{file=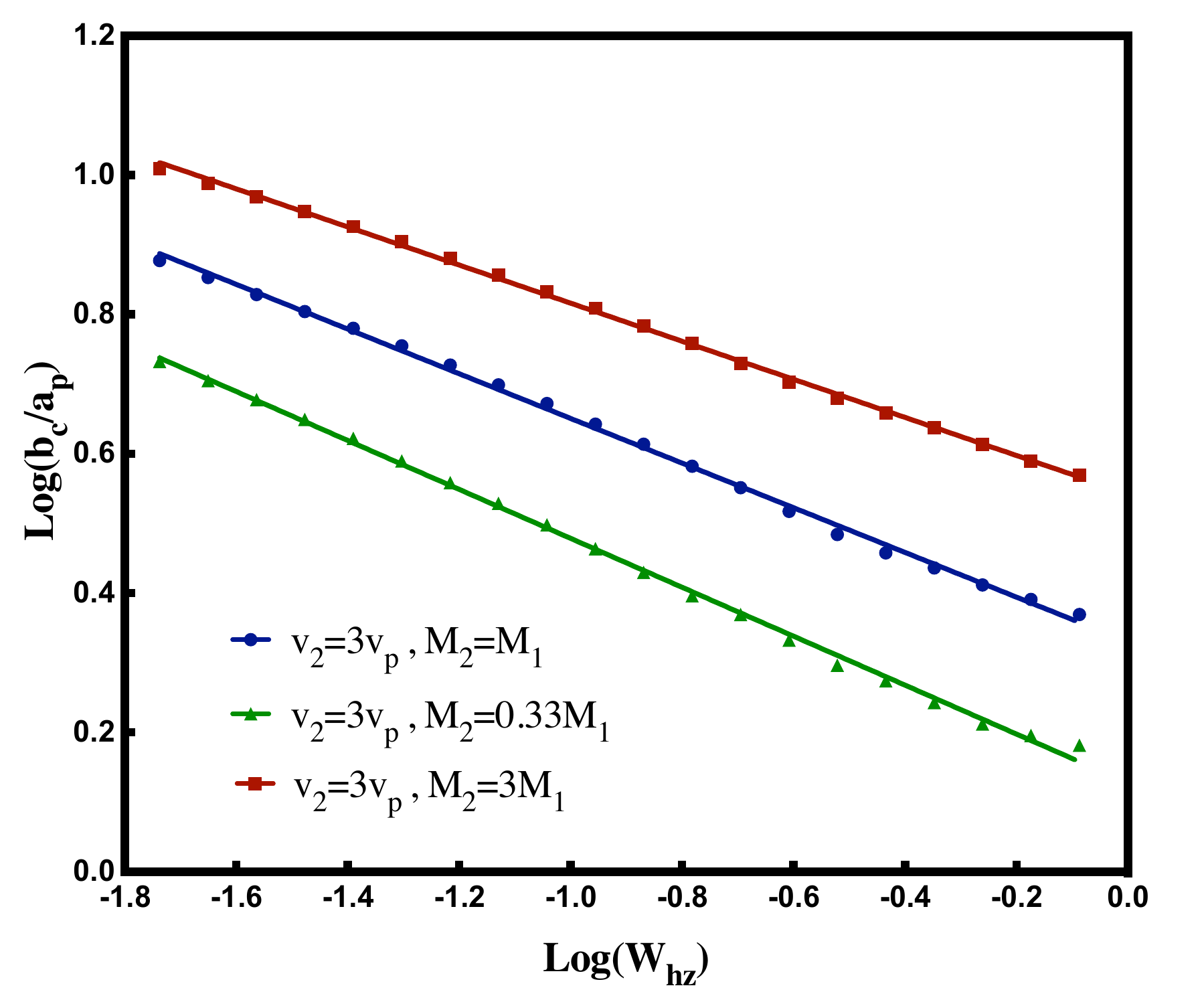,width=0.6\linewidth}}
	\vspace*{8pt}
	\caption{The critical impact parameter in logarithmic scale as a function of $W_{hz}$ in logarithmic scale with three different $M_2$ values. The two figures differ in the velocity of the encountering star ($v_2$). As shown, the relationship between critical impact parameter and $W_{hz}$ is  a power law (linear in Log-Log plot), with different powers, depending on $v_2$ and $M_2$. We interpret the dependence of critical impact parameter to $M_2$ in equation (\ref{mass-relation}).}
	\label{fig:HZ}
\end{figure}

In the previous sections, we have used a fixed habitable zone boundary. The HZ around the sun depends on the criteria on the planetary climate models. For example, Ref.~\refcite{Hart1979} calculated the inner and outer boundary to be 0.95 au and 1.01 au, respectively. Ref.~\refcite{Kasting1993} identified conservative instantaneous HZ to be between 0.95 au to 1.37 au, conservative continuous HZ to be 0.95 au to 1.15 au, and optimistic instantaneous habitable zone to be in the 0.84 au to 1.67 au, respectively. Ref.~\refcite{Kopparapu_HABITABLE_2013} also estimated the HZ boundaries to be 0.99 au and 1.70 au.
In this section, we let the habitable width be a free parameter and study the critical impact parameter in terms of the width of the 
habitable zone.

We adopt the position of the planet to be in the median radius of the habitable zone. The outcome of studying the resultant critical impact parameters versus habitable-zone width is shown in Figure (\ref{fig:HZ}) for three mass-ratios and two initial velocities of the secondary star. The scaling relation between the critical impact parameter and the width of the habitable zone is a power-law function as $ b_{d} \propto W^\beta_{hz}$ where $\beta<0$ and it is a function of the stellar mass ratio ($M_2/M_1$) and the initial velocity of the secondary star ($v_2$). We note that $W_{hz}$ depends on the stellar type and the model we select for the habitable zone. Our results show that as long as the habitable zone definitions do not differ on the scale of at least a magnitude, the rate of HZ-expelling encounters has a small dependence on the definition of the habitable zone model, compared to the environmental parameters.

\subsection{The most threatening orbital inclinations for the encounters}
\begin{figure}
	\centering
	\centerline{\psfig{file=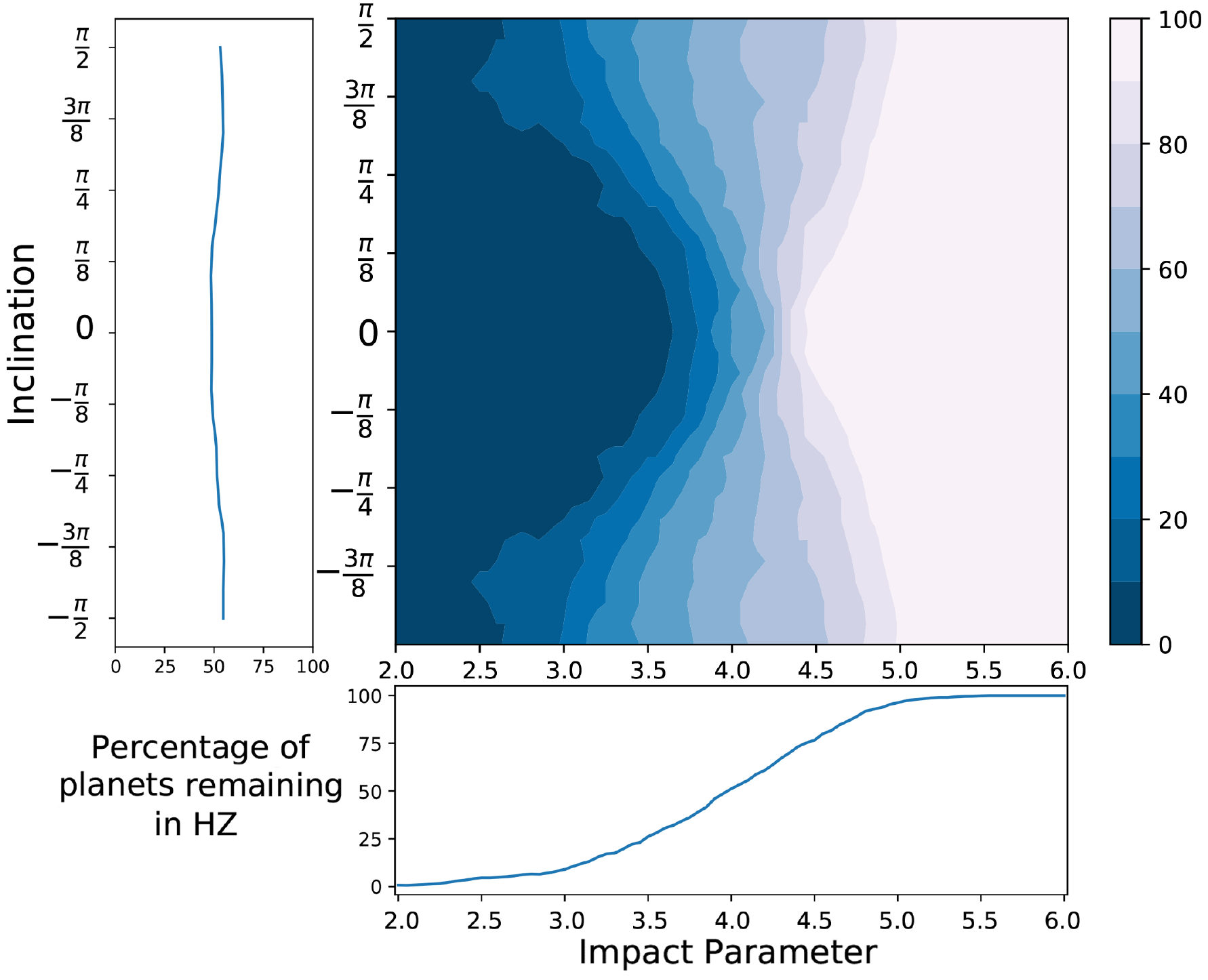,width=0.6\linewidth}}
	\caption{The ratio of events that the planet remains in the habitable zone, as a function of the inclination angle of the planetary orbital plane, and the impact parameter. For impact parameters in the range of $b/a_p<2$, a hundred percent of events are HZ expelling encounters, and for $b/a_p>6$, none of the encounters are HZ expelling. Here, we have integrated over the initial anomalies and longitude of ascending node of the planetary orbit. The HZ boundaries used are 0.8-1.2 AU.
	}
	\label{fig:inclination}
\end{figure}

Different inclinations of the planet orbit from the orbital plane of the secondary star change the ratio of HZ expelling encounters. 
In order to study the probability function for the HZ expelling encounters, we simulated planets with a fixed initial inclination angle and an impact parameter and change the initial anomaly and longitude of ascending node. We integrate over the nuisance parameters (i.e.,$\nu$ and $\Omega$) and calculate the ratio of events where the planet remains in the habitable zone for each set of inclinations and impact parameters, as shown in Figure (\ref{fig:inclination}). For a given impact parameter, a critical impact depends on the inclination angle, which for large impact parameters, larger inclination (almost parallel to the planetary orbit) are more dangerous and for the small impact parameter the smaller inclinations (almost perpendicular to the planetary orbit ) are the dangerous encounters. Averaging over the impact parameters, the percentage of planets remaining in HZ is not sensitive to the inclination angle.

\subsection{Threatening encounter rates in various stellar environments}

From evidence found in old hydrothermal vents, life on earth has been existed for about 4 billion years \cite{Dodd_Evidence_2017}. For a planet in the habitable zone of a Solar-type star to maintain life like the present Earth, we assume that the planet should be safe for at least 4 billion years.

\subsubsection{Solar neighborhood}

{\large\textbf{Data:}}We use data of the populations of stars in the solar neighborhood from Ref.~\refcite{Rickman2004}, Ref.~\refcite{Allen_Astrophysical_1973} and Ref.~\refcite{Garcia-Sanchez2001} to calculate threatening encounter rates for each stellar population as shown in Table \ref{Tab:Table1}.

\begin{table}[ph]
\tbl{Populations of stars in the solar neighborhood, with velocity dispersion (in km/s), peculiar velocity of sun (in km/s), mass (in solar mass), and stellar number density (in $10^{-3}{pc^{-3}}$). Stellar types include MK types for main-sequence stars, white dwarves (WD), and giants. (The data is adapted from Table 8 of Garcia-Sanchez (2001)\cite{Garcia-Sanchez2001})}
{\begin{tabulary}{\linewidth}{C|C|C|C|C}
	\toprule
	\textbf{Stellar type} & 
	{$v_\star~{(km/s)}~$} & {$v_\odot~{(km/s)}$~} & {$M{({M_\odot})}$} & {$n_\star{{/pc^{3}\times10^{-3}}}$}\\
	\midrule
	B0&14.7&18.6&18&0.06\\
	A0&19.7&17.1&3.2&0.27\\
	A5&23.7&13.7&2.1&0.44\\
	F0&29.1&17.1&1.7&1.42\\
	F5&36.2&17.1&1.3&0.64\\
	G0&37.4&26.4&1.1&1.52\\
	G5&39.2&23.9&0.93&2.34\\
	K0&34.1&19.8&0.78 &2.68\\
	K5&43.4&25.0&0.69 & 5.26\\
	M0&42.7&17.3&0.47 & 8.72\\
	M5&41.8&23.3&0.21 & 41.55\\
	WD&63.4&38.3&0.9 & 3.00\\
	Giants&41.0&21.0&4 & 0.4\\
	\bottomrule
\end{tabulary}	\label{Tab:Table1}}
\end{table}

{\large\textbf{Results:}} Using equation (\ref{encounter_rate}), we calculated the velocity-averaged threatening cross-section by running gravitational simulations for stellar encounters from populations present in Table \ref{Tab:Table2}.
The overall threatening encounter rate is $9.71\times10^{-5}{Gyr^{-1}}$, therefor
for a period of four billion years, the encounter probability is $3.88\times10^{-4}$. We would expect to have at least one threatening encounter out of $\approx2600$ planets in habitable orbits during the four billion years.

\begin{table}[ph]
\tbl{The threatening encounter cross-sections (in the unit of $AU^2$ and threatening encounter rate (in the unit of $Gyr^{-1}$) for the stars listed in Table \ref{Tab:Table1}.}
{\begin{tabulary}\linewidth{l|c|r}
	\toprule
	\textbf{Stellar type} & ${\langle \sigma_{T} \rangle}{(AU^{2})}$ & $\Gamma_{d} (Gyr^{-1})$ \\
	\midrule
	B0&$5.08\times10^4$&$1.74\times10^{-6}$\\
	A0&$1.16\times10^4$&$1.97\times10^{-6}$\\
	A5&$7.84\times10^3$&$2.27\times10^{-6}$\\
	F0&$5.55\times10^3$&$2.88\times10^{-6}$\\
	F5&$4.02\times10^3$&$2.48\times10^{-6}$\\
	G0&$3.47\times10^3$&$5.80\times10^{-6}$\\
	G5&$2.86\times10^3$&$7.38\times10^{-6}$\\
	K0&$2.55\times10^3$&$6.47\times10^{-6}$\\
	K5&$1.88\times10^3$&$1.19\times10^{-5}$\\
	M0&$1.35\times10^3$&$1.31\times10^{-5}$\\
	M5&$5.78\times10^2$&$2.76\times10^{-5}$\\
	WD&$1.73\times10^3$&$9.23\times10^{-6}$\\
	Giants&$9.71\times10^3$&$4.27\times10^{-6}$\\
	\midrule
	\textbf{Total} &&$9.71\times10^{-5}$ \\
	\bottomrule
\end{tabulary} \label{Tab:Table2}}
\end{table}

\subsubsection{Bulge of Milky Way galaxy}

{\large\textbf{Data:}}\cite{Wegg2013} We used the Red Clump Giants (RCGs) from the vvv survey\cite{Saito2011}  and calculated the 3-dimensional density distribution of bulge within the volume of $(\pm2.2~{\text kpc} \times \pm1.4~{\text kpc} \times \pm1.1~{\text kpc})$ surrounding the Galactic center. On the other hand,  Ref.~\refcite{Portail2015} created a dynamical model using Ref.~\refcite{Wegg2013} RCGs density measurements and velocity distribution data from Bulge Radial velocity Assay (BRAvA) spectroscopic survey \cite{Rich2006, Howard2008, Kunder2012} and evaluated the stellar mass of the inner Galactic bulge region to be $1.25-1.6 \times 10^{10} M_\odot$. From the volume of the bulge and the stellar mass, we estimate the mean stellar-mass density to be $\approx 0.5~ M_\odot/pc^{3}$.

{\large\textbf{Results:}}We repeat the same analysis for threatening encounter rates in the inner Galactic bulge. We assume the Ref.~\refcite{Kroupa2001} initial mass function for the mass function of bulge stars through the 4-billion-year period, for the mass range of $0.1 {M_\odot}$to $10 {M_\odot}$.
A differentiation from equation \ref{encounter_rate} yields:
\begin{equation}\label{differential_encounter_rate}
\frac{d{\Gamma}}{dM}= \xi(M) \langle \sigma_{T} \rangle_{(v)} v_{\star}^{bulge}~,
\end{equation}
where $\xi(M)$ is the mass function of the bulge which is normalized to the stellar-mass density in the bulge 
with $\int \xi(M) M dM = \rho_\star$,  $\langle \sigma_{T} \rangle_{(v)}$ is the threatening 
cross-section, averaged over velocity and $v_{\star}^{bulge}$=113km/s is  the velocity dispersion of stars in the inner Galactic bulge \cite{Portail2015}. We note that $\langle \sigma_{T} \rangle_{(v)}$ also depends on the mass 
of encountering stars.

\begin{figure}
	\centering
	\centerline{\psfig{file=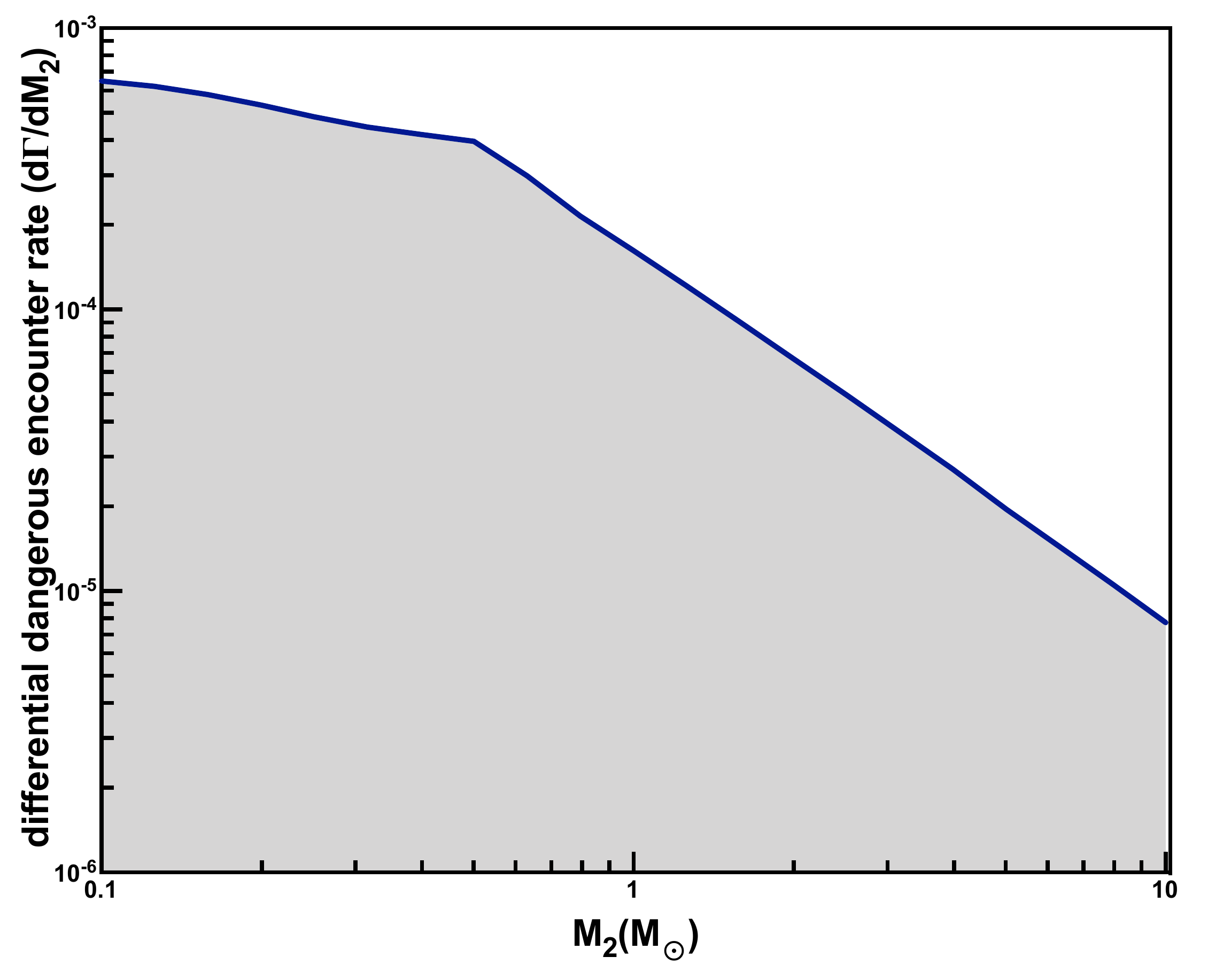,width=0.6\linewidth}}
	\caption{The differential threatening encounter rates plotted as a function of the secondary star mass in the inner region of Galactic Bulge. The integration of the differential threatening encounter rate over the secondary star mass (i.e.,the gray area) portraits the total threatening encounter rate for a planet in the region.}
	\label{fig:galactic_bulge}
\end{figure}

In the absence of velocity dispersion and number density of stars in the galactic bulge by stellar types, we used a mass-function and there is not a table of stellar types and encounter rates in comparison to the solar neighborhood. The results of differential threatening encounter rate from equation (\ref{differential_encounter_rate}) is shown in Figure (\ref{fig:galactic_bulge}). By integrating $d\Gamma/dM$ over stellar mass-function, the total threatening encounter rate is $\Gamma_{total} \approx 5.40\times 10^{-4} Gyr^{-1}$. This result shows that the catastrophe rate in the Galactic bulge region is about 6 times more than that of in the Solar neighborhood. Interpreting in life-evolution time, out of $\approx 460$ planets in the habitable zone of solar-type stars in the inner Galactic bulge region, we expect one planet to face a threatening encounter in the hypothesized 4-billion-year period required for the evolution of advanced life. 

\subsubsection{Stellar environments in the Milky Way galaxy and spheroidal dwarf galaxies}

As noted in the past sections, $\Gamma_{T}$ depends on the stellar number density, velocity distribution, and stellar mas function. To calculate and compare threatening encounter rates in various stellar environments throughout the Milky way and in the spheroidal dwarf galaxies in the neighborhood, we simulated systems in a vast parameter-space.

{\large\textbf{Data:}} The required parameters for the calculations are the {\large\textbf{Data:}} The required parameters for the calculations are the stellar number density and the velocity dispersion of stars in the environment. We assumed the same velocity distribution shape as before, and we restricted the secondary stellar mass to $1M_\odot$, to make the comparisons easier. Here we include the stellar environment data for the Milky Way globular clusters, and Milky Way dwarf spheroidals (dSph).

The data for the Milky Way GCs come from Ref.~\refcite{Baumgardt2018a}. They acquired kinematic and structural parameters of 112 Milky Way GCs by fitting N-body simulations. We used the stellar mass-density in the cluster cores, with the velocity dispersions. We calculated the approximate envelope of the cluster parameters and used it to show the region illustrating the overall GC populations.

The velocity dispersion and stellar number density distribution for dSph satellite galaxies are taken from Ref.~\refcite{Walker2007}. Here, the full range of velocity dispersion and number density of stars is assumed and we have not taken into account the possible relations between the distribution of the two parameters.

\begin{figure*}
	\centering
	\centerline{\psfig{file=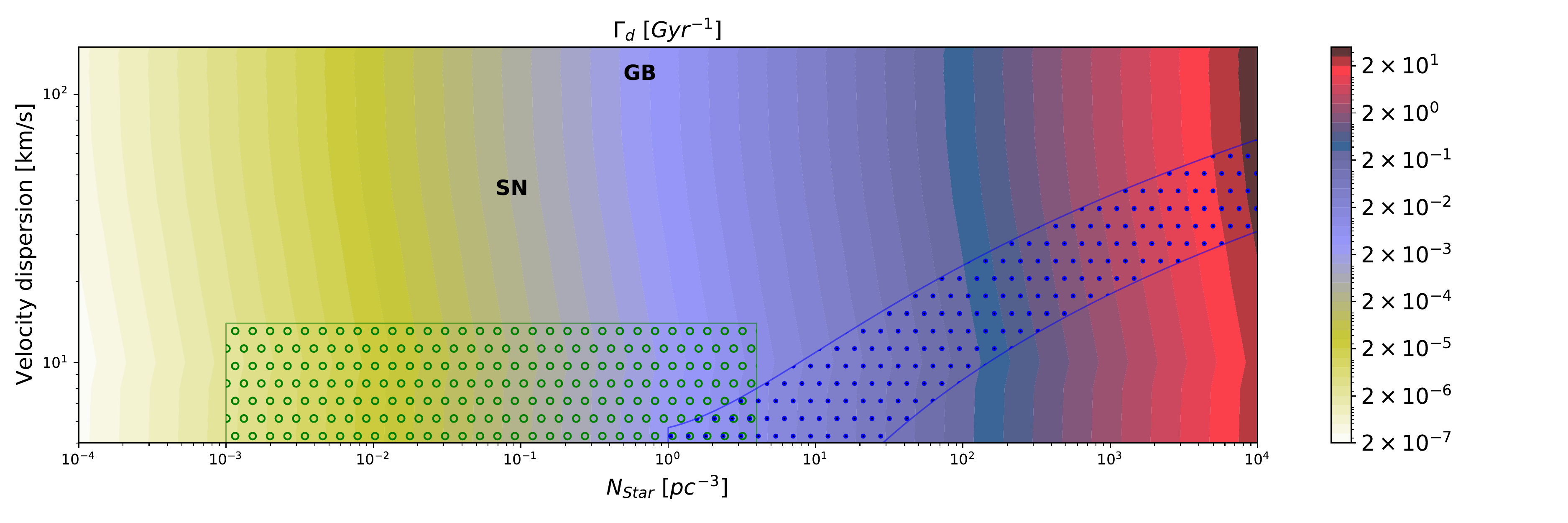,width=\linewidth}}
	\caption{General threatening encounter rate ($\Gamma_{T}$), for general parameter spaces of stellar number density and velocity dispersion. The solar neighborhood position is shown by \textbf{SN}, and galactic bulge is shown by \textbf{GB}. The blue area with filled small dots illustrates the globular clusters, and the green rectangular area with circles shows the spheroidal dwarf galaxies surrounding Milky way. Stellar birth-clusters possess much lower velocity dispersion than the lower limits of this figure.}
	\label{fig:general_population}
\end{figure*}

{\large\textbf{Results:}} Figure \ref{fig:general_population} shows the result and the threatening encounter rates for the stellar environments. For open clusters, the velocity dispersion is much lower than the galactic neighborhood (less than 10km/s), therefore in the encounter rate calculations, we can neglect the rates for OC members, making them just as safe as the surrounding space Ref.~\refcite{Soubiran2018}. Note that the encounter rates are independent of velocity dispersion changes in high velocities. In the velocities around 10km/s, one can notice a turnaround; the threatening encounter rate increases with the decrease of velocity dispersion for $V_\star<10km/s$. This behavior can be explained by the changeover seen in figure \ref{fig:velocity}, wherein the high-speed encounters, $b_d \propto v^{-0.5}$. If we assume a similar behavior for the threatening impact rate, then $\sigma_{T} \propto v^{-1}$, and therefore according to equation \ref{encounter_rate}, $\Gamma_{T}$ becomes independent of velocity. In low velocities, $b_d \propto v^{-1}$, and with the same arguments as above, $\Gamma_{T} \propto v^{-1}$. In the middle, the threatening impact parameter log-log slope becomes small, and therefor, $\Gamma_{T} \propto v^Q$, where $0<Q<1$, explaining the proportionality seen in Figure \ref{fig:general_population}.

\section{Discussions}
\label{Discussions}

In this work, we studied the close stellar encounters that can disrupt the orbit of habitable planets and deprive the planets of habitability conditions. We took the parent star with the planet as a binary system while the encountering star plays the role of the third gravitating object. Using the numerical calculation for the gravitational interaction, we also considered the gravitational effect of the planet on the parent star and encountering star. In this study, we had a six-dimension parameter-space for describing this encounter and studied the dependencies on the initial parameter. The parameter space in our study is divided two (i) geometric and (ii) dynamic parts. The geometric part consisted of the inclination angle (i.e.,$i$), orbital ascending node (i.e $\Omega$), and encountering impact parameter (i.e.,$b$) of the planet. The dynamical parameter space consisted of the velocity and the mass of the encountering star and width of the habitable zone. We adopted the parent star and the planet similar to the Sun-Earth system and defined a critical impact parameter of which closer encounters can displace the planet from its habitable zone.

We studied the dependency of critical impact-parameter on the velocity and mass of encountering stars. We demonstrated that high-speed encounters ($v_2 \gg v_p$) have smaller critical impact parameters with the proportionality of $b_d \propto  v_2^{-0.5}$. 
The relation becomes ($b_d \propto v_2^{-1}$) for slower encounters, as a result of encounter perigee of the orbit dependency with the velocity of the star in the initial distant position. The stellar encounter velocities comparable to planetary velocity are less dependent on the stellar velocity.
In case of fast encounters, $b_c$ is comparable to the asteroid belt orbital radii. For encounters with the speed of encountering star comparable to the speed of the planet (i.e $v_2 \sim v_p$), critical impact parameters are about the radius of Jupiter orbit, while $b_c$ values for low-speed encounter ($v_2 \ll v_p$) are as large as Uranus orbital radius. Also for the stellar masses, our analysis showed a power-law relationship of the critical impact-parameter with relative stellar masses of the encountering star and the parent star ($b_d\propto \left( M_2/M_1\right)^{0.34}$).  Note that the discussed velocity-$b_c$ relations for slow and fast stellar velocities, and the mass-$b_c$ relation are not dependent of the habitable zone boundaries and mass of the star and the planet, as long as the planetary mass in negligible. Therefore the relation can be implemented for systems with different HZ boundaries and parent star masses.

Studying the dependency of critical impact-parameter with habitable zone width ($W_{hz}$), a power-law relationship is observed while the power depends on the velocity of the secondary star ($v_2$) and the stellar mass-ratio ($M_2/M_1$). While we used Kasting HZ, in order to investigate close stellar encounters using other HZ definitions and boundaries, this relationship can play an important role in generalizing the current results. 

In order to measure the encounter rates that can cause an orbital change habitability hazard, we defined the threatening encounter rate ($\Gamma_T$). We reported the results for our calculations for $\Gamma_{T}$ in different stellar environments. Our results showed that the threatening encounter rate for a planet in the habitable zone around a Solar-type star is $\approx 1.07\times 10^{-4} Gyr^{-1}$. Therefore in the solar neighborhood, we expect one planet out of $\approx 2250$ to experience a threatening encounter in a period of 4 billion years, or in other words, the chances of experiencing such encounters is $\approx0.044\%$ in 4 billion years, or $0.1\%$ in 9 billion years (comparable to the lifetime of a solar-type star).

We also studied threatening encounter rates in the inner Galactic bulge region of Milky Way galaxy. Utilizing the number density, mass-function, and velocity-dispersion of stars in the inner Galactic bulge, we estimated the $\Gamma_{T}\approx 5.97\times 10^{-4} Gyr^{-1}$. In other words, in $\approx 400$ stars in the region, one experiences the threatening stellar encounter in the 4-billion-year period. Therefor, the catastrophe rate is roughly 5.5 times higher in the Galactic bulge compared to the solar neighborhood. 

We extended our parameter space survey to calculate the threatening encounter rate for a wide set of stellar environments and included the Milky Way globular clusters and nearby spheroidal dwarf galaxies' stellar number density and velocity dispersions, shown in Fig \ref{fig:general_population}. The number density of stars played the most important role in the encounter rates for the Galactic regions. It is remarkable that the variations of threatening encounter rates due to velocity dispersion is not considerable, therefore almost in any Galactic region, $\Gamma_{T}$ follows the variations of stellar number density.

\section{Conclusions}
\label{Conclusions}

Amid stellar encounter hazards such as expulsion of planets and Oort cloud disruption, threatening stellar encounters able to change initially habitable planetary orbits out of the circumstellar habitable zone are shown to be important in continuous habitable planets. In this work we have shown that in regions like the solar neighborhood, the probability for an initially habitable planet to get out of the HZ as a result of threatening stellar encounters is as small as $0.044\%$ in 4 billion years. In Galactic bulge region the encounter rates are higher, but the rates are only significant in the most dense stellar regions such as globular clusters. Also the encounter rates were shown to be not strongly dependent on the velocity dispersion of the stellar region. This work has specifically studied single planetary systems, therefore the study has been on direct gravitational effects of encountering stars. 


\section*{Acknowledgements}

This research was supported by Sharif University of Technology Office of vice President for Research under grant no. G950214.

\bibliographystyle{ws-ijmpd}
\bibliography{Bojnordi_ArXiv_2021a}

\begin{thebibliography}{10}

\bibitem{NASA-ea}
{ {NASA Exoplanet Archive}}.

\bibitem{Huang_The_1959}
S.-S. Huang, {\em Publ. Astron. Soc. Pacific} {\bf 71}  (1959) 421.

\bibitem{Huang_The_1960}
S.-S. Huang, {\em Publ. Astron. Soc. Pacific} {\bf 72}  (1960) 489.

\bibitem{Kasting1993}
J.~F. Kasting, D.~P. Whitmire and R.~T. Reynolds, {\em Icarus} {\bf 101}
  (1993) 108.

\bibitem{Kopparapu_HABITABLE_2013}
R.~Kopparapu, R.~Ramirez, J.~F. Kasting, V.~Eymet, T.~D. Robinson,
  S.~Mahadevan, R.~C. Terrien, D.-G. Shawn, V.~Meadows and R.~Deshpande, {\em
  Astrophys. J.} {\bf 765}  (2013)   131.

\bibitem{Wignall2001}
P.~B. Wignall, {\em Earth Sci. Rev.} {\bf 53}  (2001) 1.

\bibitem{Alvarez2003}
W.~Alvarez, {\em Astrobiology} {\bf 3}  (2003) 153.

\bibitem{Pandey1987}
O.~P. Pandey and J.~G. Negi, {\em Geophys. J. R. Astron. Soc.} {\bf 89}  (1987)
  857.

\bibitem{Abbott2002}
D.~H. Abbott and A.~E. Isley, {\em Earth Planet. Sci. Lett.} {\bf 205}  (2002)
  53.

\bibitem{Lyttleton1965}
R.~A. Lyttleton and S.~Yabushita, {\em Mon. Not. R. Astron. Soc.} {\bf 129}
  (1965) 105.

\bibitem{Rasio1996}
F.~A. Rasio and E.~B. Ford, {\em Science (80-. ).} {\bf 274} (Nov 1996) 954.

\bibitem{Luhman2005}
K.~L. Luhman, P.~D. Alessio, N.~Calvet, L.~Hartmann, S.~T. Megeath and G.~G.
  Fazio, {\em Astrophys. J.}   (2005) 93.

\bibitem{Hurley2002}
J.~R. Hurley and M.~M. Shara, {\em The Astrophysical Journal} {\bf 1}  (2002)
  1, \href{http://arxiv.org/abs/0108350}{{\ttfamily arXiv:0108350}}.

\bibitem{Veras2012}
D.~M. Veras and S.~N. Raymond, {\em Mon. Not. R. Astron. Soc. Lett.} {\bf 421}
  (2012) 117.

\bibitem{Kobayashi2001}
H.~Kobayashi and S.~Ida, {\em Icarus} {\bf 153}  (2001) 416.

\bibitem{vincke2015}
K.~Vincke, A.~Breslau and S.~Pfalzner, {\em Astron. Astrophys.} {\bf 577}
  (2015)   A115.

\bibitem{Clarke_Star_1991}
C.~J. Clarke and J.~E. Pringle, {\em Mon. Not. R. Astron. Soc.} {\bf 249}
  (1991) 584.

\bibitem{Hall_Energetics_1996}
S.~M. Hall, C.~J. Clarke and J.~E. Pringle, {\em Mon. Not. R. Astron. Soc.}
  {\bf 278}  (1996) 303.

\bibitem{Ostriker_Capture_1994}
E.~C. Ostriker, {\em Astrophys. J.} {\bf 424}  (1994) 292.

\bibitem{Lestrade2011}
J.-F. Lestrade, E.~Morey, A.~Lassus and N.~Phou, {\em Astron. Astrophys.} {\bf
  532}  (2011)   A120.

\bibitem{Li2015}
G.~Li and F.~C. Adams, {\em Mon. Not. R. Astron. Soc.} {\bf 448}  (2015) 344.

\bibitem{Malmberg_The_2011}
D.~Malmberg, M.~B. Davies and D.~C. Heggie, {\em Mon. Not. R. Astron. Soc.}
  {\bf 411}  (2011) 859.

\bibitem{Jimenez-Torres2011}
J.~J. Jim\'enez-Torres, B.~Pichardo, G.~Lake and H.~Throop, {\em Mon. Not. R.
  Astron. Soc.} {\bf 418}  (2011) 1272.

\bibitem{Marzari2013}
F.~Marzari and G.~Picogna, {\em Astron. Astrophys.} {\bf 550}  (2013)   A64.

\bibitem{Picogna2014}
G.~Picogna and F.~Marzari, {\em Astron. Astrophys.} {\bf 564}  (2014)   A28.

\bibitem{Zwart_The_2015}
S.~F. {Portegies Zwart} and L.~J{\'{i}}lkov{\'{a}}, {\em Mon. Not. R. Astron.
  Soc.} {\bf 451}  (2015) 144.

\bibitem{Fragner2009}
M.~M. Fragner and R.~P. Nelson, {\em Astron. Astrophys.} {\bf 505}  (2009) 873.

\bibitem{Craig2013}
J.~Craig and M.~R. Krumholz, {\em Astrophys. J.} {\bf 769}  (2013)   150.

\bibitem{Chapman1994}
C.~R. Chapman and D.~Morrison, {\em Nature} {\bf 367}  (1994) 33.

\bibitem{Oort1950}
J.~H. Oort, {\em Bull. Astron. Institutes Netherlands} {\bf 11}  (1950) 91.

\bibitem{Rickman2008}
H.~Rickman, M.~Fouchard, C.~Froeschl{\'{e}} and G.~B. Valsecchi, {\em Celest.
  Mech. Dyn. Astron.} {\bf 102}  (2008) 111,
  \href{http://arxiv.org/abs/0804.2560}{{\ttfamily arXiv:0804.2560}}.

\bibitem{Hills1981}
J.~G. Hills, {\em Astrophys. J.} {\bf 86}  (1981) 1730.

\bibitem{Matese2002}
J.~J. Matese and J.~J. Lissauer, {\em Icarus} {\bf 157}  (2002) 228.

\bibitem{Wickramasinghe2008}
J.~T. Wickramasinghe and W.~M. Napier, {\em Mon. Not. R. Astron. Soc.} {\bf
  387}  (2008) 153.

\bibitem{Bobylev2017}
V.~Bobylev, {\em arXiv.org} {\bf 61}  (2017) 883.

\bibitem{Bailer2017}
C.~A. Bailer-Jones, {\em Proc. Int. Astron. Union} {\bf 12}  (2017) 144.

\bibitem{Rybizki2018}
C.~A. Bailer-Jones, J.~Rybizki, R.~Andrae and M.~Fouesneau, {\em Astron.
  Astrophys.} {\bf 616}  (2018).

\bibitem{Rampino2015}
M.~R. Rampino, {\em Mon. Not. R. Astron. Soc.} {\bf 448}  (2015) 1816.

\bibitem{Mazeeva2004}
O.~A. Mazeeva, {\em Sol. Syst. Res.} {\bf 38}  (2004) 325.

\bibitem{Jakubik2008}
M.~Jakub{\'{i}}k and L.~Neslu{\v{s}}an, {\em Contrib. Astron. Obs. Skaln.
  Pleso} {\bf 38}  (2008) 33.

\bibitem{Feng2014}
F.~Feng and C.~A. Bailer-Jones, {\em Mon. Not. R. Astron. Soc.} {\bf 442}
  (2014) 3653.

\bibitem{Martinezbarbosa2016}
C.~A. Martinez-Barbosa, L.~Jilkova, S.~{Portegies Zwart} and A.~G.~A. Brown,
  {\em Mon. Not. R. Astronmical Soc.}   (2016).

\bibitem{Cai2017}
M.~X. Cai, M.~B.~N. Kouwenhoven, S.~F.~P. Zwart and R.~Spurzem, {\em Mon. Not.
  R. Astron. Soc.} {\bf 1706}  (2017)   arXiv:1706.03789,
  \href{http://arxiv.org/abs/1706.03789}{{\ttfamily arXiv:1706.03789}}.

\bibitem{VanElteren2019a}
A.~van Elteren, S.~P. Zwart, I.~Pelupessy, M.~X. Cai, S.~L. McMillan,
  S.~{Portegies Zwart}, I.~Pelupessy, M.~X. Cai and S.~L. McMillan, {\em
  Astron. Astrophys.} {\bf 624}  (2019) 1,
  \href{http://arxiv.org/abs/1902.04652}{{\ttfamily arXiv:1902.04652}}.

\bibitem{Garcia-Sanchez2001}
J.~Garc{\'{i}}a-S{\'{a}}nchez, P.~R. Weissman, R.~A. Preston, D.~L. Jones,
  J.-F.~F. Lestrade, D.~W. Latham, R.~P. Stefanik and J.~M. Paredes, {\em
  Astron. Astrophys.} {\bf 379}  (2001) 634.

\bibitem{Rein_REBOUND_2012}
H.~Rein and S.-F. Liu, {\em Astron. Astrophys.}   (2011).

\bibitem{Rein2014a}
H.~Rein and D.~S. Spiegel, {\em Mon. Not. R. Astron. Soc.} {\bf 446}  (2014)
  1424.

\bibitem{Hut1985}
P.~Hut and S.~Tremaine,  {\bf 90}  (1985) 1548.

\bibitem{Bobylev2014}
V.~V. Bobylev and A.~T. Bajkova, {\em Mon. Not. R. Astron. Soc.} {\bf 441}
  (2014) 142.

\bibitem{Hills1984}
J.~G. Hills, {\em Astron. J.} {\bf 89}  (1984) 1559.

\bibitem{Rickman1976}
H.~Rickman, {\em Publ. House Czechoslov. Acad. Sci.}   (1976).

\bibitem{Hart1979}
M.~H. Hart, {\em Icarus} {\bf 37}  (1979) 351.

\bibitem{Dodd_Evidence_2017}
M.~S. Dodd, D.~Papineau, T.~Grenne, J.~F. Slack, M.~Rittner, F.~Pirajno,
  O.~Jonathan and C.~T.~S. Little, {\em Nature} {\bf 543}  (2017) 60.

\bibitem{Rickman2004}
H.~Rickman, C.~Froeschl{\'{e}}, C.~Froeschl{\'{e}} and G.~B. Valsecchi, {\em
  Astron. Astrophys.} {\bf 428}  (2004) 673.

\bibitem{Allen_Astrophysical_1973}
C.~W. Allen, {\em {Astrophysical quantities}}, 3rd edn. (Atholone Press, 1973).

\bibitem{Wegg2013}
C.~Wegg and O.~Gerhard, {\em Mon. Not. R. Astron. Soc.} {\bf 435}  (2013) 1874.

\bibitem{Saito2011}
R.~K. Saito, M.~Hempel, D.~Minniti, P.~W. Lucas, M.~Rejkuba, I.~Toledo, O.~A.
  Gonzalez, J.~Alonso-Garcia, M.~J. Irwin, E.~Gonzalez-Solares, S.~T. Hodgkin,
  J.~R. Lewis, N.~Cross, V.~D. Ivanov, E.~Kerins, J.~P. Emerson, M.~Soto, E.~B.
  Amores, S.~Gurovich, I.~Dekany, R.~Angeloni, J.~C. Beamin, M.~Catelan,
  N.~Padilla, M.~Zoccali, P.~Pietrukowicz, C.~M. Bidin, F.~Mauro, D.~Geisler,
  S.~L. Folkes, S.~E. Sale, J.~Borissova, R.~Kurtev, A.~V. Ahumada, M.~V.
  Alonso, A.~Adamson, J.~I. Arias, R.~M. Bandyopadhyay, R.~H. Barba, B.~Barbuy,
  G.~L. Baume, L.~R. Bedin, A.~Bellini, R.~Benjamin, E.~Bica, C.~Bonatto,
  L.~Bronfman, G.~Carraro, A.~N. Chene, J.~J. Claria, J.~R.~A. Clarke,
  C.~Contreras, A.~Corvillon, R.~de~Grijs, B.~Dias, J.~E. Drew, C.~Farina,
  C.~Feinstein, E.~Fernandez-Lajus, R.~C. Gamen, W.~Gieren, B.~Goldman,
  C.~Gonzalez-Fernandez, R.~J.~J. Grand, G.~Gunthardt, N.~C. Hambly, M.~M.
  Hanson, K.~G. Helminiak, M.~G. Hoare, L.~Huckvale, A.~Jordan, K.~Kinemuchi,
  A.~Longmore, M.~Lopez-Corredoira, T.~Maccarone, D.~Majaess, E.~L. Martin,
  N.~Masetti, R.~E. Mennickent, I.~F. Mirabel, L.~Monaco, L.~Morelli, V.~Motta,
  T.~Palma, M.~C. Parisi, Q.~Parker, F.~Penaloza, G.~Pietrzynski, G.~Pignata,
  B.~Popescu, M.~A. Read, A.~Rojas, A.~Roman-Lopes, M.~T. Ruiz, I.~Saviane,
  M.~R. Schreiber, A.~C. Schroder, S.~Sharma, M.~D. Smith, L.~Sodre, J.~Stead,
  A.~W. Stephens, M.~Tamura, C.~Tappert, M.~A. Thompson, E.~Valenti, L.~Vanzi,
  N.~A. Walton, W.~Weidmann and A.~Zijlstra, {\em Astron. Astrophys.} {\bf 537}
  (Nov 2011).

\bibitem{Portail2015}
M.~Portail, C.~Wegg, O.~Gerhard and I.~Martinez-Valpuesta, {\em Mon. Not. R.
  Astron. Soc.} {\bf 448}  (2015) 713.

\bibitem{Rich2006}
R.~M. Rich, D.~B. Reitzel, C.~D. Howard and H.~Zhao, {\em Astrophys. J.} {\bf
  688} (Nov 2006) 1060.

\bibitem{Howard2008}
C.~D. Howard, R.~M. Rich, D.~B. Reitzel, A.~Koch, R.~{De Propris} and H.~Zhao,
  {\em Astrophys. J.} {\bf 688}  (2008) 1060.

\bibitem{Kunder2012}
A.~Kunder, A.~Koch, R.~{Michael Rich}, R.~{De Propris}, C.~D. Howard, S.~A.
  Stubbs, C.~I. Johnson, J.~Shen, Y.~Wang, A.~C. Robin, J.~Kormendy, M.~Soto,
  P.~Frinchaboy, D.~B. Reitzel, H.~Zhao and L.~Origlia, {\em Astron. J.} {\bf
  143}  (2012).

\bibitem{Kroupa2001}
P.~Kroupa, {\em Mon. Not. R. Astron. Soc.} {\bf 322}  (2001) 231.

\bibitem{Baumgardt2018a}
H.~Baumgardt and M.~Hilker, {\em Mon. Not. R. Astron. Soc.} {\bf 478}  (2018)
  1520, \href{http://arxiv.org/abs/1804.08359}{{\ttfamily arXiv:1804.08359}}.

\bibitem{Walker2007}
M.~Walker, M.~Mateo, E.~Olszewski, O.~Gnedin, X.~Wang, B.~Sen and M.~Woodroofe,
  {\em Astrophys. J.} {\bf 667}  (2007) 53.

\bibitem{Soubiran2018}
C.~Soubiran, T.~Cantat-Gaudin, M.~Romero-G{\'{o}}mez, L.~Casamiquela, C.~Jordi,
  A.~Vallenari, T.~Antoja, L.~Balaguer-N{\'{u}}{\~{n}}ez, D.~Bossini,
  A.~Bragaglia, R.~Carrera, A.~Castro-Ginard, F.~Figueras, U.~Heiter, D.~Katz,
  A.~Krone-Martins, J.~F. {Le Campion}, A.~Moitinho and R.~Sordo, {\em Astron.
  Astrophys.} {\bf 619}  (2018) 1,
  \href{http://arxiv.org/abs/1808.01613}{{\ttfamily arXiv:1808.01613}}.

\end{thebibliography}

\end{document}